# Magnetic resonance imaging of flow and mass transfer in electrohydrodynamic liquid bridges.

Adam D. Wexler[a,*], Sandra Drusova[a], Elmar C. Fuchs[a], Jakob Woisetschläger[b], Gert Reiter[c], Michael Fuchsjäger[d], and Ursula Reiter[d]

[a]*Applied Water Physics, Wetsus European Center of Excellence for Sustainable Water Technology, 8911MA Leeuwarden, Netherlands*

[b]*Institute for Thermal Turbomachinery and Machine Dynamics, Working Group Metrology - Laser Optical Metrology, Technical University of Graz, 8010 Graz, Austria*

[c]*Siemens AG Healthcare, 8054 Graz, Austria*

[d]*Division of General Radiology, Department of Radiology, Medical University of Graz, 8036Graz, Austria*

* corresponding author: adam.wexler@wetsus.nl +31 (0)58 284 3013 http://www.wetsus.eu

**Abstract**

Here we report on the feasibility and use of magnetic resonance imaging based methods to the study of electrohydrodynamic (EHD) liquid bridges. High speed tomographic recordings through the longitudinal axis of water bridges were used to characterize the mass transfer dynamics, mixing, and flow structure. By filling one beaker with heavy water and the other with light water it was possible to track the spread of the proton signal throughout the total liquid volume. The mixing kinetics are different depending on where the light nuclei are located and proceeds faster when the anolyte is light water. Distinct flow and mixing regions are identified in the fluid volumes and it is shown that the EHD flow at the electrodes can be counteracted by the density difference between water isotopes. MR phase contrast imaging reveals that within the bridge section two separate counter propagating flows pass one above the other in the bridge.

**1. Introduction**

Electrohydrodynamic (EHD) liquid bridges are a peculiar phenomenon which has recently come into focus as useful for studying the intersection of continuum and molecular scale interactions in liquid matter. EHD bridges are easy to produce, requiring only a low conductivity polar liquid, two reservoirs, and a high potential low current DC power source (Wexler et al. 2014). These bridges can be formed by a number of polar liquids including water (Woisetschläger et al. 2011), and in the latter case are referred to as floating water bridges (Fuchs et al. 2007). They possess complex flow patterns which have previously been studied using optical methods such as schlieren visualization (Sammer et al. 2015) and high speed imaging (Fuchs et al. 2007). These methods have revealed bi-directional flow which is corroborated by mass transfer observations and for water results in a net mass transport from anode to cathode (Woisetschläger et al. 2010). The gross flow dynamics can be understood within an EHD framework as the system acts as extended Taylor pump (Melcher and Taylor 1969). However, this only applies to the whole system flow and is not sufficient to describe the flow in the bridge section itself (Burcham and Saville 2002). The departure from theory comes as the structure of the bridge appears to be radially heterogeneous with a core region where flow is along the axis of the bridge and an outer skin which supports a spiral flow behavior. Polyimide tracer particles confined to the outer skin of the bridges were measured using laser Doppler anemometry (LDA) and yielded a tangential velocity of 30 cm/s with an axial velocity of ±20 cm/s, however it was unclear whether this motion was due to the fluid motion or charging of the tracer particles in the high electric field (Woisetschläger et al. 2010). Inelastic ultraviolet light scattering supports the radial structure of the bridge by showing that physical quantities (e.g. density, velocity of sound, or kinematic viscosity) associated with molecular mobility exhibit a radial gradient (Fuchs et al. 2009; Fuchs et al. 2011).

It was thus intriguing to consider the possibility of using magnetic resonance imaging (MRI) to add information about floating water bridges by virtue of generating tomographic images throughout the sample volume. The recovered signal could provide insight into proton density, magnetic relaxation times or velocities within the sample (Bernstein et al. 2004). There is precedent for the use of magnetic resonance velocimetry in turbulent mixing (Benson et al. 2009), swirling (Grundmann et al. 2012), structured (Wassermann et al. 2014), and even cytoplasmic flows (Van De Meent et al. 2010). This work is a first look at testing the feasibility as well as understanding the challenges associated with magnetic resonance based methods where high intensity electric fields are present in the scanner. Furthermore, MR specifically permits the visualization of how flows composed from different isotopes interact, something not possible using any other method. The aim of the present study was thus to investigate mass transfer and velocity fields in operating EHD bridges employing MRI – an approach which to the authors' best knowledge has not previously been attempted.

## 2. Methods

2.1 EHD bridges in an MRI

EHD bridges were prepared in a manner consistent with the method described in (Wexler et al. 2014) and modified to be compatible with the magnetic environment in a medical MRI scanner (Magnetom Sonata, Siemens Healthcare GmBH, Erlangen, Germany). Non-magnetic materials were used in the construction of the fixating armature, cable supports, and electrode mounts. The beaker spouts were left in contact throughout the measurements to reduce the chance of bridge rupture, liquid leakage, and electrical arcing – all of which could pose a hazard to the MRI scanner. Though this produced a bridge system without a free-hanging section these "zero-length" bridges (see figure 1) behaved in a manner consistent with extended bridges. The seven EHD bridges used in this study, four for isotope mixture measurements and three for flow investigations, were operated in a voltage limiting regime between 15.00-18.00 kV and open currents between 600-1600 μA. The higher current than previously reported is due to the larger diameter of the bridge ~6-8 mm on account of the zero-length. The experimental set-up, illustrated in figure 2, was placed within a standard circular-polarized knee coil, and aligned so that the long axis of the bridge was parallel to the main magnetic field $B_0$. The bridge center was located at the isocenter in the head-feet and right-left directions, and several millimeters above the isocenter in anterior-posterior direction.

The bridges were always arranged with the anode placed towards the foot end of the patient table and the cathode towards the head end. A plastic box filled with 500 mL agar and covered with a glass plate formed the electrically insulating support base for the beakers. The agar phantom was used as a signal reference for intensity normalization. All materials were fixed in place using MR compatible tape which does not produce artifacts in the recorded images. Platinum foil (99.999% Pt, MaTeck GmbH, Jülich, Germany) was soldered to both the high voltage (HV) and ground leads. These wires were sufficiently long (~10m) to reach from the experiment at the MR scanner isocenter to the power supply (HCP 30000-300, FuG Elektronic GmbH, Schechen, Germany) located just outside the room via an RF suppressing pass-through in the Faraday cage wall. A low voltage resonant RLC tank circuit installed on the HV coaxial shield and ground wires shifted most of the induced radio frequency interference (RFI) bands outside the operational frequency range of the scanner, however one band persisted. Mass flow was tracked by filling the one beaker with heavy water (99.9 % $D_2O$, DLM-4-100, Cambridge Isotope Laboratories, Inc., Andover, MA, USA) and the other with an equal volume of freshly prepared ultrapure light water (~99.98 % $H_2O$, Cat No. ZMQSP0D01, Millipore Corp., MA, USA). The starting mass used in these experiments was 61.3±0.7 g light water, 65.5±2.3 g heavy water corresponding to a starting volume of ~63 mL in each beaker. The variability in mass was due to the criterion that the liquid levels at the spouts be equal. Heavy water produces no signal as the precession frequency of the deuterium (D) nucleus at 1.5 T is outside the operating bandwidth of the MRI scanner (Graessner 2013) thus it is possible to track the mass transfer and mixing rates of protium (H) nuclei in an operating bridge as a function of both intensity and location (Wang et al. 2013). In all experiments the starting conductivity of the water was below 1 $\mu S \cdot cm^{-1}$. MR flow imaging used equal volumes of light water (67.0±0.5 g) in both beakers.

2.2 MRI sequences and protocols

Isotope mixture and flow experiments were started with the acquisition of an isotropic 3-dimensional (3D) spoiled fast low-angle shot (FLASH) sequence (Bernstein et al. 2004) covering the experimental set-up. Protocol parameters were as follows: Echo time (TE), 3.5 ms; repetition time (TR), 9.3 ms; flip angle, 5°; resolution 0.7x0.7x0.7 mm$^3$; bandwidth, 130 Hz/pixel; field-of-view (FOV), 140x140 mm$^2$; number of slices, 88. The sagittal (or almost sagittal, in case of small experimental misalignment) plane containing the exact long axis of the bridge was reconstructed from the 3D data set by multiplanar reformatting (see figure 3) and was used as central imaging plane for further measurements. This sampling volume reveals all salient features of an operating EHD bridge system – i.e. anolyte, catholyte, and bridge.

Isotope mixture was visualized by repeated acquisition of a 3D FLASH measurement with protocol parameters: TE, 1.9 ms; TR, 4.5 ms; flip angle, 5°; resolution 1.3x1.3x5.0 mm$^3$; bandwidth, 300 Hz/pixel; FOV, 160x160 mm$^2$; number of slices, 16. The acquisition time of 3 s per 3D slab determined the time resolution of the imaged mixture process.

For flow measurements a 2D FLASH-based phase contrast sequence with three-directional velocity encoding by a simple four-point velocity encoding scheme (Bernstein et al. 2004) was employed. Velocity encoding (VENC) was set to 70 cm/s in all directions. Further protocol parameters were as follows: TE, 5.1 ms; TR, 41 ms; flip angle, 10°; resolution 0.5x0.8x4.0 mm$^3$; bandwidth, 225 Hz/pixel; FOV, 140x140 mm$^2$; number of averages, 19. The bridge volume was covered with 7-9 parallel overlapping slices (slice distance of 2 mm) in (almost) sagittal image orientation determined by the planning procedure described above.

2.3 Image analysis

To study isotope mixture, the time evolution of mean signal intensities for anolyte (SI$_{anolyte}$), catholyte (SI$_{catholyte}$) and agar phantom (SI$_{reference}$) were derived in the central sagittal slice employing standard MR software (syngo.MR, Siemens Healthcare GmBH, Erlangen, Germany). The regions of interest (ROI) in the anolyte ($ROI_{anolyte}$) and catholyte ($ROI_{catholyte}$) comprised the respective fluid volumes and were manually adapted in each time frame; the ROI in the agar phantom ($ROI_{reference}$) was kept fixed (figure 4). Relative volumes of anolyte ($relVol_{anolyte}$) and catholyte ($relVol_{catholyte}$) at a time *t* were estimated according to

$$\mathrm{relVol_i(t)} = \frac{\mathrm{ROI_i(t)}}{\mathrm{ROI_i(t_0)}}, \tag{1}$$

with *i* = anolyte or catholyte and $t_0$ the time of bridge ignition. Relative total fluid volume ($relVol_{total}$) at a time *t* was determined as

$$\mathrm{relVol_{total}(t)} = \frac{\mathrm{ROI_{anolyte}(t)} + \mathrm{ROI_{catholyte}(t)}}{\mathrm{ROI_{anolyte}(t_0)} + \mathrm{ROI_{catholyte}(t_0)}}. \tag{2}$$

As signal intensity in FLASH sequence is proportional to proton density (Bernstein et al. 2004), relative signal intensities (relSI) of the anolyte $relSI_{anolyte} = SI_{anolyte}/(SI_{anolyte}+SI_{catholyte})$ and catholyte $relSI_{catholyte} = SI_{catholyte}/(SI_{anolyte}+SI_{catholyte})$ were employed as measures of proton densities in respective beakers. By choosing low flip angle and short echo time, the FLASH signal intensity depended only moderately on changes in T1 and T2* relaxation times, which might be caused by chemical exchange of hydrogen isotopes and/or heating during mixture (Bernstein et al. 2004). The change of relative total signal intensity for anolyte and catholyte with respect to the reference signal $SI_{reference}$ is defined as

$$\mathrm{relSI_{total}(t)} = \frac{\mathrm{SI_{anolyte}(t) \cdot ROI_{anolyte}(t)} + \mathrm{SI_{catholyte}(t) \cdot ROI_{catholyte}(t)}}{\mathrm{ROI_{anolyte}(t)} + \mathrm{ROI_{catholyte}(t)}} / \mathrm{SI_{reference}(t)} \tag{3}$$

and was used to check for possible relaxation time changes during the experiment.

Phase contrast images of light water flow experiments were evaluated by means of prototype software (4D Flow, Siemens Healthcare GmBH, Erlangen, Germany) allowing visualization and analysis of 3D velocity fields (Reiter et al. 2013).

## 3. Results and Discussion

3.1 Electrical polarization, bridge ignition and noise

Prior to the existence of a bridge the experiment can be understood as a simple dielectric capacitor. As the applied voltage is ramped electrical energy is stored within the liquid and other dielectric materials present. These materials become polarized and will leak charge into the surrounding environment. The DC power is converted to material fluctuations in the Hz to kHz range which generates an audible sound commonly encountered in capacitor charging circuits. Additionally, the system will leak and spray charge in the form of ion or corona wind by which air is charged at sharp edges or points in the setup, this plasma will generate broadband radio frequency interference (RFI) the intensity of which is maximum in the seconds preceding and during bridge ignition. Image acquisition in the isotope mixture experiments was begun prior to the application of electrical energy (t=0 s) thus it is possible to observe the charging of the system in the radio frequency domain as shown in figure 5. The voltage ramping rate was between 800-1000 V/s with bridge ignition occurring between 10-15 kV (t=12-15 s). The impulsive stochastic noise which covers the entire image frame becomes spatially constrained (t=21-24 s) after the bridge is established. This indicates that the interfering signal is now confined to a narrower frequency band introducing zipper-like artifacts parallel to phase encoding direction (Weishaupt et al. 2008). Between subsequent frames the artifact showed a slow drift movement indicating that the extraneous frequency is time varying. In case of flow experiments, image slices with zipper-like artifacts close to the bridge were repeatedly measured.

3.2 Mass transport and isotope mixing

Two experimental cases were prepared to track mass transport:

>Case 1 heavy water anolyte and light water catholyte;

>Case 2 light water anolyte and heavy water catholyte.

Both cases were investigated twice, however, in only one instance of each case did the system completely mix as evidenced visually and by converged relSI values. The reason for this is not readily apparent and did not correlate with the variations in starting mass.

A summary sequence of representative images at regular time intervals from the two experimental cases is shown in figure 6. A movie compiled from the MR images stack of a single instance for case 1 is provided in the supplemental materials. In both cases the bright signal from $^1$H labeled molecules are transported across the bridge. The light water floats on top of the heavier deuterated water as expected given the density difference between isotopes. As the experiment progresses the light water spreads throughout both beakers. The protium signal moves slowly downward into the originally deuterated volume. At the same time the signal intensity in both beakers falls. Most of the mixing is observed to occur in the upper fluid volume and along the bridge itself but is not especially strong in the vicinity of the electrodes. Schlieren flow visualization (Sammer et al. 2015) has shown that a downward flow is present at both electrodes, however, in the case of the heavy water volume this flow is insufficient to overcome the density gradient resulting in an upward buoyancy force that repels lower density flows of $H_2O$ and HDO. Heating in the bridge section will further enhance any buoyancy forces in the system. Thus, an unmixed volume of heavy water remains at the bottom of the respective beaker. In case 1 (heavy water anolyte, figure 6a) this unmixed layer persists for nearly the entire experiment (~1700 s), whereas in case 2 (heavy water catholyte, figure 6b) mixing is nearly complete within 900 seconds. The duration of the experiments is listed in table 1 and it can be seen that in general case 1 required much longer to mix than case 2.

**Table 1** Comparison of relative changes in mass and temperature for the two cases considered. The starting mass was 61.3±0.7 g light water and 65.5±2.3 g heavy water. All bridges lost some mass during operation. The starting temperature for all liquids was 23°C.

| Case | Experimental Duration [s] | Relative mass change anolyte [g] | Relative mass change catholyte [g] | Relative mass change total [g] | Relative temperature change anolyte [°C] | Relative temperature change catholyte [°C] |
|---|---|---|---|---|---|---|
| 1 ($D_2O$ Anolyte) | 1794 ±5 | -22±2 | 10.5±1.1 | -11.5±1.1 | 31.3±0.5 | 34.8±0.2 |
| 2 ($D_2O$ Catholyte) | 1029±109 | -17.0±0.1 | 7.5±1.8 | -7.5±1.8 | 24.3±2.4 | 27.1±1.6 |

In addition dynamics in the relative volume are visible as changes in the respective liquid levels of each beaker. In case 1 (figure 6a) the initial net mass flow is towards the anode after some time this flow reverses and the catholyte volume increases significantly. Volume only flows from anode to cathode in case 2 (figure 6b). For simplicity we will use the convention throughout the paper that forward flow is from the anode to the cathode, and reverse flow from the cathode towards the anode. Forward flow the typical behavior for water bridges prepared with a uniform isotope composition.

At the conclusion of the MR measurements the final weight of the liquid in each reservoir along with the temperature was recorded and is compared in table 1. The deviation values are the standard error of the mean (SEM). In all bridges measured some mass was lost, this is presumed due to evaporation during the measurement time caused by Ohmic heating and electrospray (Smith et al. 2002). The average final solution temperature was 52.3±1.6°C. This indicates operation temperatures higher than typically encountered in normal water bridges and may be due in part to energy liberated by isotope mixing. The temperatures in the catholyte were consistently higher than the anolyte for both cases.

The time courses of relSI (upper panels) and relVol (lower panels) for both anolyte and catholyte were measured using the ROIs illustrated in figure 4. Data show that in one instance of each case the relSI converge to the same value and these systems are called 'fully mixed' (figure 7a, 7b) as compared to 'partially mixed' (figure 7c, 7d) where the values have not yet converge. It is assumed that given enough time all systems will become fully mixed. Calculations based on the final relVol$_{total}$ values confirms that material is lost in the experiments as shown in table 2 and figure 7 (lower panels). The discrepancy between volume and mass loss measurements is likely related to reduced density at the elevated temperatures found in these bridges. The agreement is nonetheless within a few percent and is given in table 2. The duration of the reverse flow from case 1 is taken to be the elapsed time before the anolyte and catholyte volumes again become equivalent. Similar times were recorded for the fully mixed case 873 s (figure 7b) and the partially mixed case 900 s (figure 7d), although in the latter the cross-over is less well pronounced. The bridge system regardless of case eventually reaches a steady state where the relative volumes are ~0.75-0.8 and ~1.2-1.25 for the anolyte and catholyte, respectively. The change of relative signal intensity in all four experiments (table 2) initially proceeds at approximately the same rate, and requires on average 378±8 s to reach a level where change in relSI has reached half of the fully mixed equilibrium value.

**Table 2** Comparison of ROI based measurements for the two cases in both the fully and partially mixed states. The relative loss of volume, and signal intensity are compared to the time to half mixed.

| Initial D2O Volume | State of Mixing | relative volume loss | Time to halfway mixed state [s] | relSI$_{total}$ loss |
|---|---|---|---|---|
| Anolyte | full | 2.2% | 382 | 62% |
| | partial | 5.1% | 399 | 62% |
| Catholyte | full | 5.0% | 355 | 55% |
| | partial | 3.4% | 376 | 58% |

Measurement of the local proton density is central to following the mixing dynamics. Figure 8 shows the time courses of the relative total signal intensities of anolyte and catholyte in the case of heavy water anolyte and light water catholyte and in the case of light water anolyte and heavy water catholyte, respectively. The decrease of $relSI_{total}$ is monotonic and reaches similar final values in all cases. This indicates changes of signal intensities due to changes in T1 and/or T2*, which in turn should be caused by both, chemical exchange of hydrogen isotopes (Wang et al. 2013) and/or substantial heating during mixture (Narten 1964; Quesson et al. 2000) as indicated in table 1. This substantially decreases proportionality between signal intensity and proton density.

3.3. Flow in the bridge

The flow direction in an EHD liquid bridge is a dynamic phenomenon and best understood as the balance of EHD and hydrostatic forces (Widom et al. 2009; Marín and Lohse 2010; Woisetschläger et al. 2012). The steady state flow direction is material dependent on account of the stability and transport kinetics of ions in the liquid (Woisetschläger et al. 2012). For water this results in preferential flow from anolyte to catholyte. However, periodic flow reversal is an established trait of EHD bridges and is thought to result from the hydrostatic pressure temporarily overcoming EHD transport. Likewise, reverse flow can occur immediately following bridge ignition and is likely due to brief force imbalances associated with bridge ignition (Woisetschläger et al. 2010). These types of reverse flow are transient and last no more than a few to tens of seconds in bridges made using identical solutions in both beakers.

The reverse flow observed in the isotope mixing experiments is thus an unusual situation. It can be explained as an additional displacement generated by the isotope density difference. That the reverse flow persists for many tens of minutes indicates separate flows of different density ($H_2O$, HDO, $D_2O$) within the system. This will retard local mixing and prolong the time require for the system to equilibrate as is seen to be the case. Density gradients are thus a deterministic factor in the dynamic mass balance of a water bridge.

In order to better visualize the flow dynamics and mixing forces in the bridge section 3D velocity maps were recorded for three light water bridges (figure 9a). Despite the limitation of averaging non-stationary flow for several minutes of data acquisition the results for the three bridges were consistent and provide a first look at the complex flow dynamics inside the bridge. Previously flow has only been observed optically in EHD liquid bridges using a number of techniques (Fuchs et al. 2007; Fuchs 2008; Fuchs et al. 2008) that suffer from limitations due to the variable cylindrical geometry.

Two counter current flows are present in the bridge (figure 9b), one flowing over the other as they cross from one beaker to the other. The mean peak velocities ($v_f$ forward direction or $v_r$ reverse) of these flows were sampled at two positions in the bridge -- halfway between the bridge base and center. Forward flow shall always refer to flow from anode to cathode, and reverse flow the opposite. Mean peak velocities found in the slice half-way between anode base and bridge center (figure 9c) were $v_f$ = 18 ± 5 cm/s and $v_r$ = 27 ± 6 cm/s, respectively. At half-way between catholyte base and bridge center (figure 9d) $v_f$ = 22 ± 2 cm/s and $v_r$ = 26 ± 4 cm/s. The deviation values are the standard error of the mean (SEM). These peak velocities in the forward and reverse direction are on the same order of magnitude as those measured for the longitudinal flow using LDA. However, due to the limited spatial resolution of MR phase contrast images it is as yet difficult to determine whether or not the spiral flow observed previously in the outer bridge is a genuine feature present only in a very thin surface shear layer (Woisetschläger et al. 2012). It can be extrapolated from these measurements that in the normal steady state condition where the forward flow velocity is less than the reverse velocity the cross-sectional area of the forward flow must be greater than that of the reverse flow in order to maintain the equilibrium. Indeed the shape of water bridges is known to be amphora like (Morawetz 2012), and furthermore swings in the equilibrium height difference between anolyte and catholyte is associated with bridge diameter fluctuations. Such a relationship is consistent with the Bernoulli flow model discussed in (Widom et al. 2009; Marín and Lohse 2010; Woisetschläger et al. 2012).

It is also interesting to consider the action of the magnetic field on the liquid volume. The configuration of the bridge system was chosen in part to minimize the Lorentz force that would act on the bridge during operation. As a quick check of the maximum forces expected on the system we can quickly calculate the Lorentz force $F$ acting on a current carrying wire:

$$F = Il \times \boldsymbol{B_0} = \|Il\|\|\boldsymbol{B_0}\| \sin \theta \qquad (4)$$

Where $I$ is the current over the bridge (1.6 mA), $l$ is the length (<1 cm), and $\boldsymbol{B}_0$ the magnetic field strength (1.5 T). As the cross-product is sensitive to the relative orientation of the current path and the magnetic field lines in a purely coaxial situation, which is approximately the case here, the net force will be zero as the $sin(0) = 0$; as a worst case scenario one could imagine current conduction perpendicular to $\boldsymbol{B_0}$, $sin(90) = 1$, and thus F becomes for the values reported in this study

$$F = \|(1.6 \cdot 10^{-3} A)(10^{-2} m)\|\|1.5T\| \sin 90 = 2.4 \cdot 10^{-5} N = 2.4 \cdot 10^{-4} \frac{g}{cm \cdot s} \qquad (5)$$

This value is rather small when compared to the EHD forces responsible for the bridge which are capable of lifting and transporting several grams of water per second over barrier heights in excess of 1 cm. Thus, by several orders of magnitude we can see that the observed behavior is likely not the result of magnetic field interactions with charges in the liquid volume and is rather a genuine feature of interacting isotope labelled EHD flows. For a more complete discussion on the role that electromagnetic forces play on polarizable fluid elements see the work of Engel and Friedrichs (2002).

## 4. Conclusions

Despite the challenges and potential hazards of investigating EHD bridges using MRI the method is viable and provides useful information on the internal workings of not only the bridge section but the entire fluid system. As amendment to previous observations of preferential transport from anode to cathode in EHD water bridges the measurements with heavy water anolyte and light water catholyte show that net mass transport proceeds in the opposite direction despite relatively rapid mixing. Once the chemical mixing of isotopes proceeds beyond a critical value ($relSI_{anolyte} \approx 0.32$) and the dominant molecular species becomes HDO normal flow behavior returns. MR phase contrast imaging of the vector fields in the bridge section shows distinct counter-propagating flows. Both convective transport and diffusion can be tracked but not disambiguated using the methods employed in this study. It is necessary to better understand how isotope composition, temperature, electric fields, and sequence design influence the recovered signal. Thus far MRI has proven a useful tool in helping to understand the internal structure of EHD mass flow and nuclear transport under the influence of moderate electric field gradients.

**Acknowledgements**

This work was performed in the cooperation framework of Wetsus, European Center of Excellence for Sustainable Water Technology (www.wetsus.eu). Wetsus is co-funded by the Dutch Ministry of Economic Affairs and Ministry of Infrastructure and Environment, the Province of Fryslân, and the Northern Netherlands Provinces. ADW, SD, ECF, and JW wish to thank the participants of the research theme Applied Water Physics for the fruitful discussions and their financial support.

8:e82212. doi: 10.1371/journal.pone.0082212

Sammer M, Wexler A, Kuntke P, et al (2015) Proton production, neutralisation and reduction in a floating water bridge.

Smith JN, Flagan RC, Beauchamp JL (2002) Droplet evaporation and discharge dynamics in electrospray ionization. J Phys Chem A 106:9957–9967. doi: 10.1021/jp025723e

Van De Meent JW, Sederman AJ, Gladden LF, Goldstein RE (2010) Measurement of Cytoplasmic Streaming in Single Plant Cells by Magnetic Resonance Velocimetry. J Fluid Mech 642:5. doi: 10.1017/S0022112009992187

Wang FN, Peng SL, Lu CT, et al (2013) Water signal attenuation by D2O infusion as a novel contrast mechanism for 1H perfusion MRI. NMR Biomed 26:692–698. doi: 10.1002/nbm.2914

Wassermann F, Loosmann F, Egger H, et al (2014) Flow through Tetradecahedrons. 7–10.

Weishaupt D, Köchli VD, Marincek B (2008) How does MRI work?: An Introduction to the Physics and Function of Magnetic Resonance Imaging. Springer Science & Business Media

Wexler AD, López Sáenz M, Schreer O, et al (2014) The preparation of electrohydrodynamic bridges from polar dielectric liquids. J Vis Exp e51819. doi: 10.3791/51819

Widom a., Swain J, Silverberg J, et al (2009) Theory of the Maxwell pressure tensor and the tension in a water bridge. Phys Rev E 80:016301. doi: 10.1103/PhysRevE.80.016301

Woisetschläger J, Gatterer K, Fuchs EC (2010) Experiments in a floating water bridge. Exp Fluids 48:121–131. doi: 10.1007/s00348-009-0718-2

Woisetschläger J, Wexler AD, Holler G, et al (2011) Horizontal bridges in polar dielectric liquids. Exp Fluids 52:193–205. doi: 10.1007/s00348-011-1216-x

Woisetschläger J, Wexler AD, Holler G, et al (2012) Horizontal bridges in polar dielectric liquids. Exp Fluids 52:193–205.
17

**Figures**

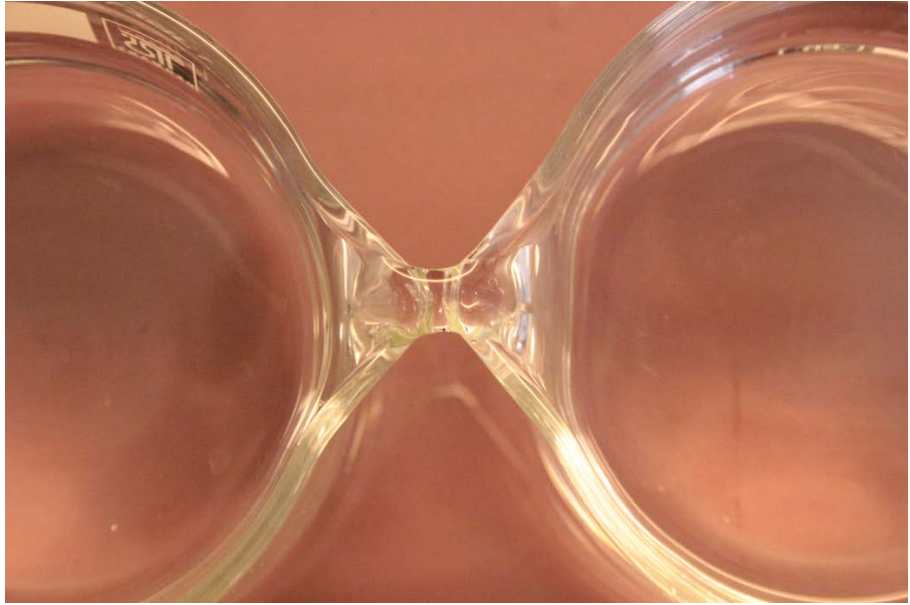

**Fig. 1** Zero-distance EHD bridge prepared with ultrapure light water. The bridge waist diameter in this image is 6mm

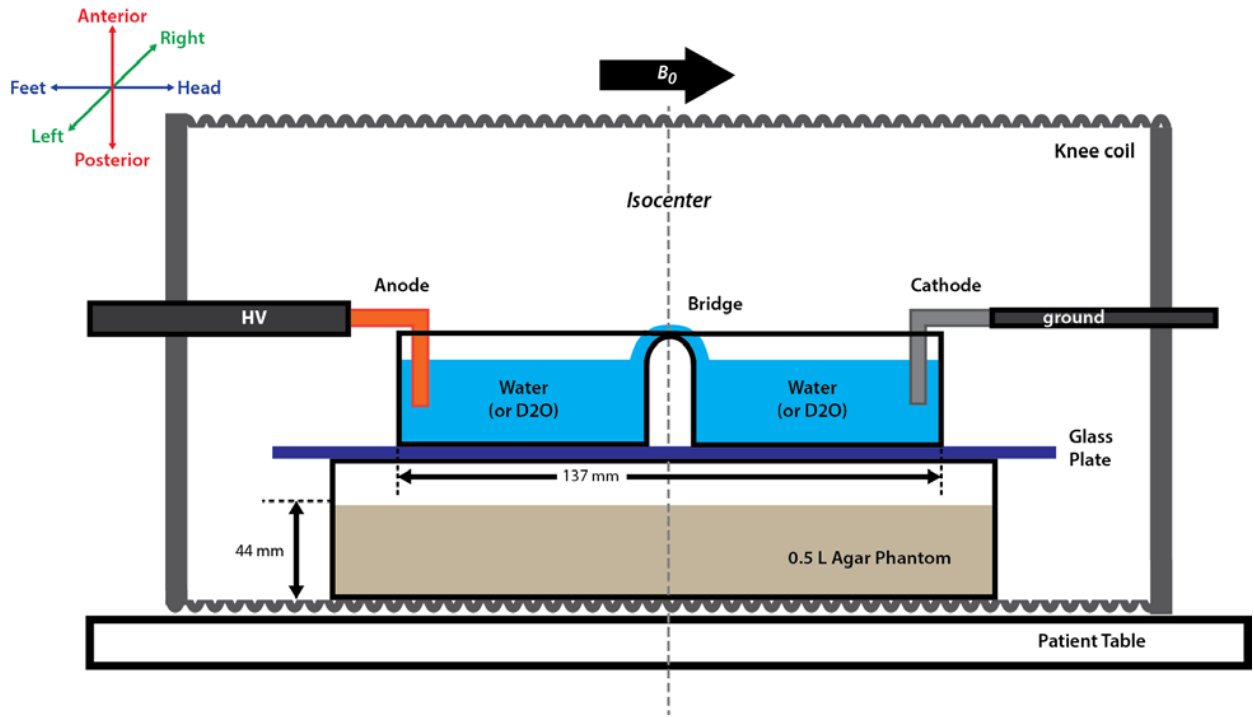

**Fig. 2** Diagram of the experimental setup used for the presented study. Main orthogonal directions (feet-head, anterior-posterior and left-right) are indicated in the upper left corner

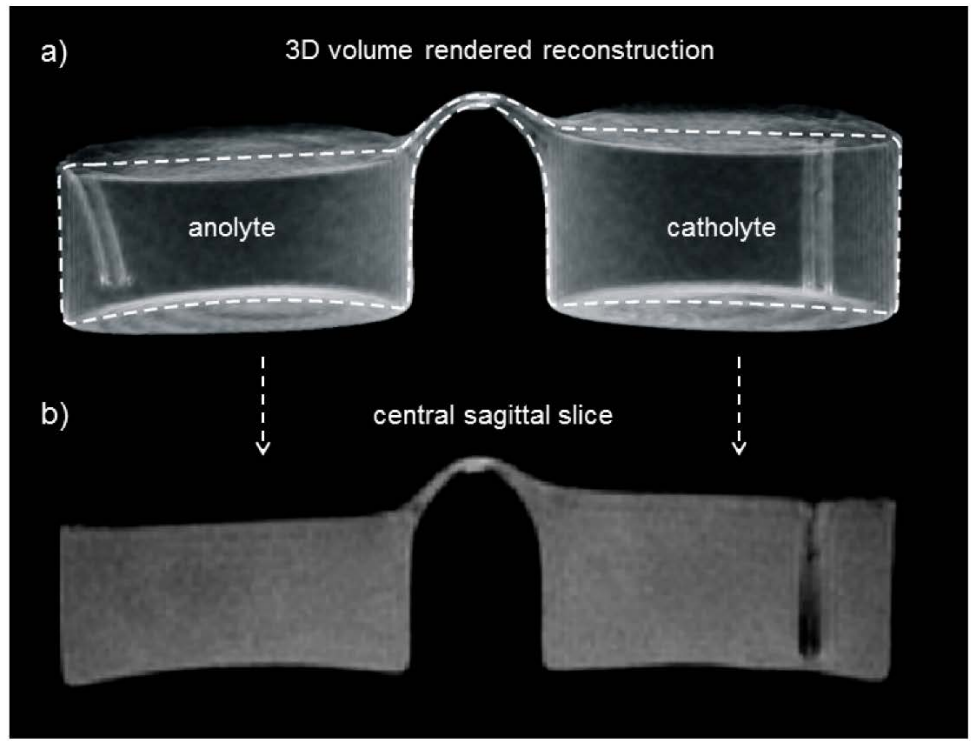

**Fig. 3** Tomographic volume rendering of a typical water bridge (a) and the multiplanar reformatted central sagittal slice (b) used in this study

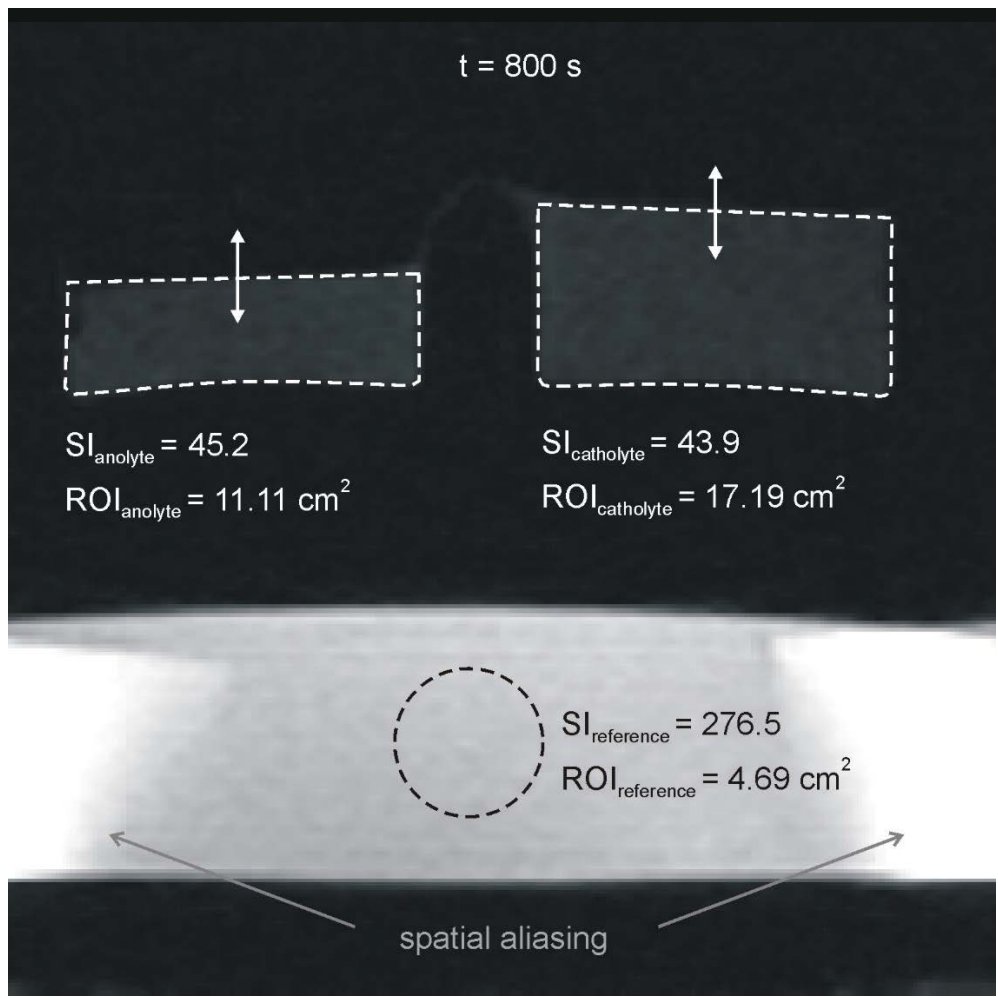

**Fig. 4** Definition of regions of interest in the isotope mixture experiments. Fluid levels of anolyte and catholyte (here at t = 800 s) were adapted in each time frame (as indicated by arrows). The bridge was excluded from analysis

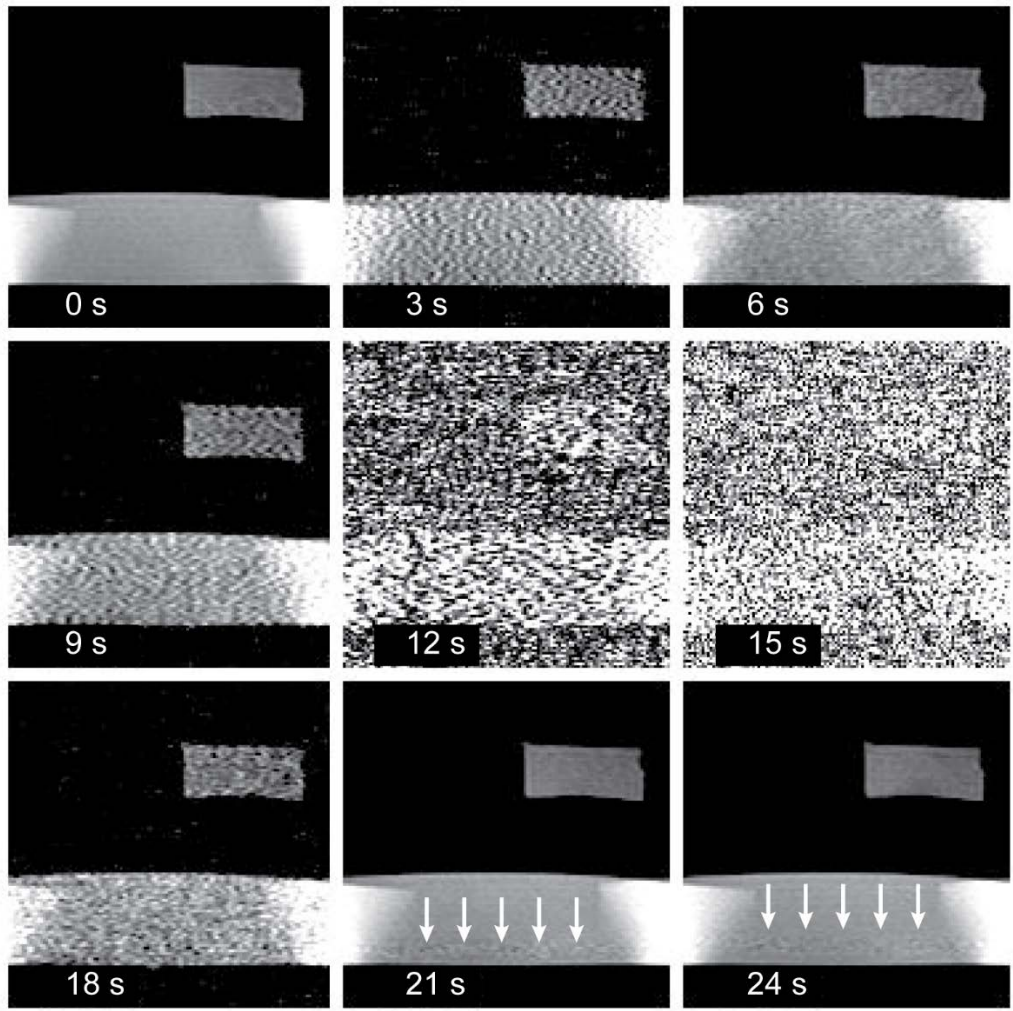

**Fig. 5** Noise from radio frequency interference imaged during bridge charging, ignition, and operation. Impulse noise is visible beginning shortly after the application of high voltage (t = 3 s). As the voltage increases the noise intensity builds and reduces image contrast. Bridge ignition (t = 12-15 s) is accompanied by a brief electromagnetic pulse that obscures the field of view. Starting at t = 21 s a zipper-like artifact is visible parallel to the phase encoding direction.

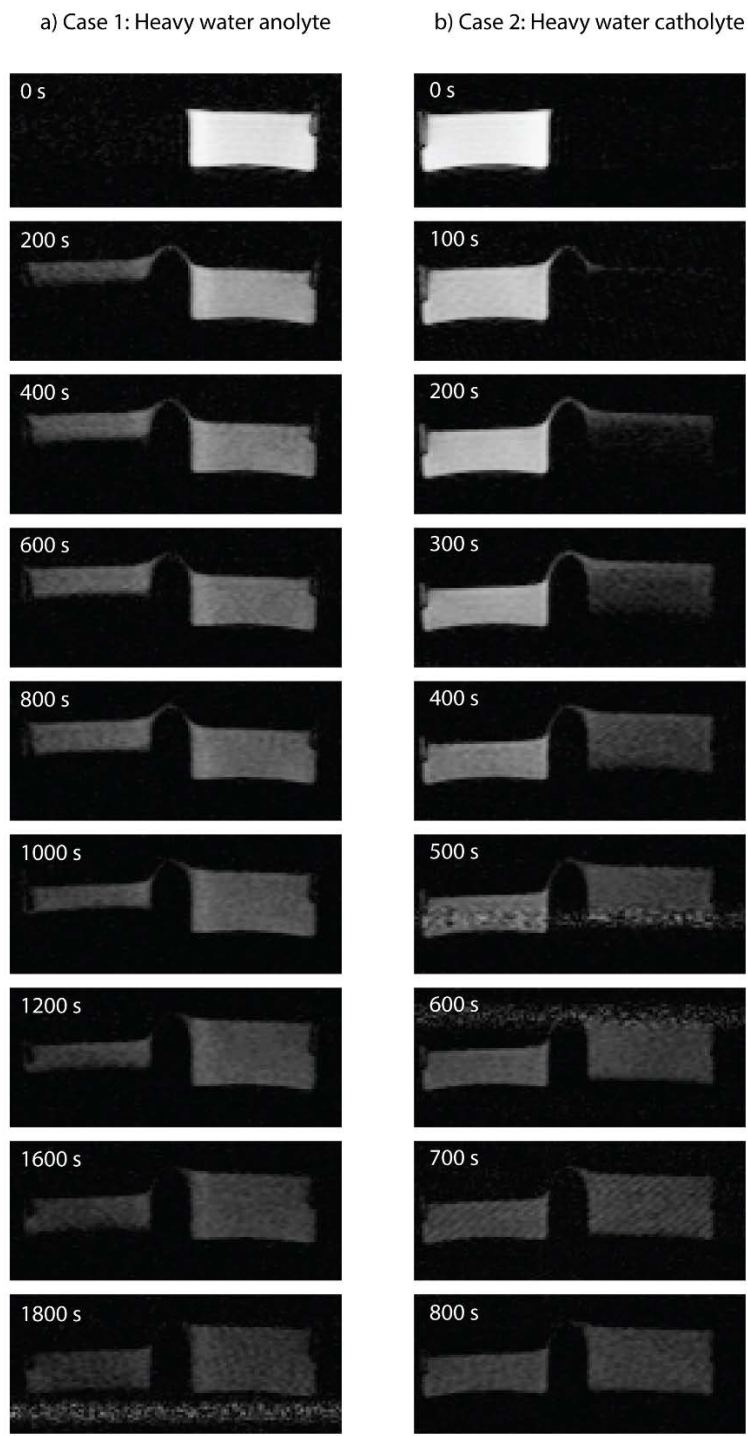

**Fig. 6** Representative image sequence showing the transport of $^1$H nuclei (brighter signal) in the bridge beginning with light water in the cathode (panel a) or anode (panel b) beakers. The transport and mixing of light water with heavy water is much faster in the case where the anolyte is the $^1$H source, requiring half as much time and producing fully mixed volumes. Zipper-like artifact can be seen moving through the measurement frames. The agar signal phantom is not shown for clarity.

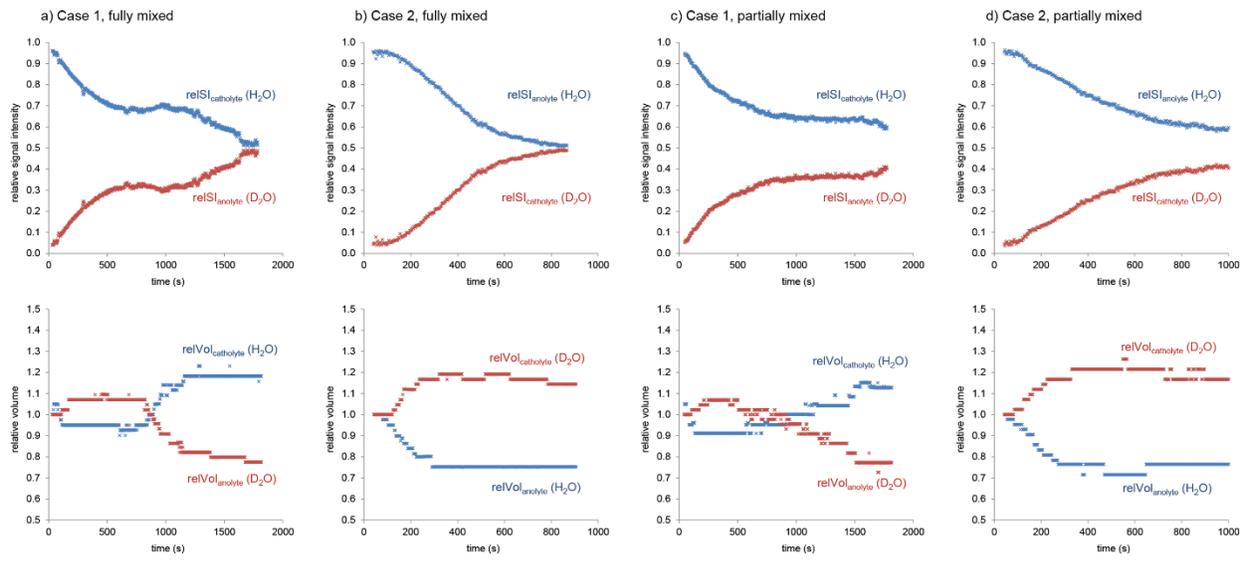

**Fig. 7** Time courses of relative signal intensities of the anolyte and catholyte (upper panels) as well as their relative volumes (lower panels). Results from both the fully mixed (7a, 7b) and partially mixed

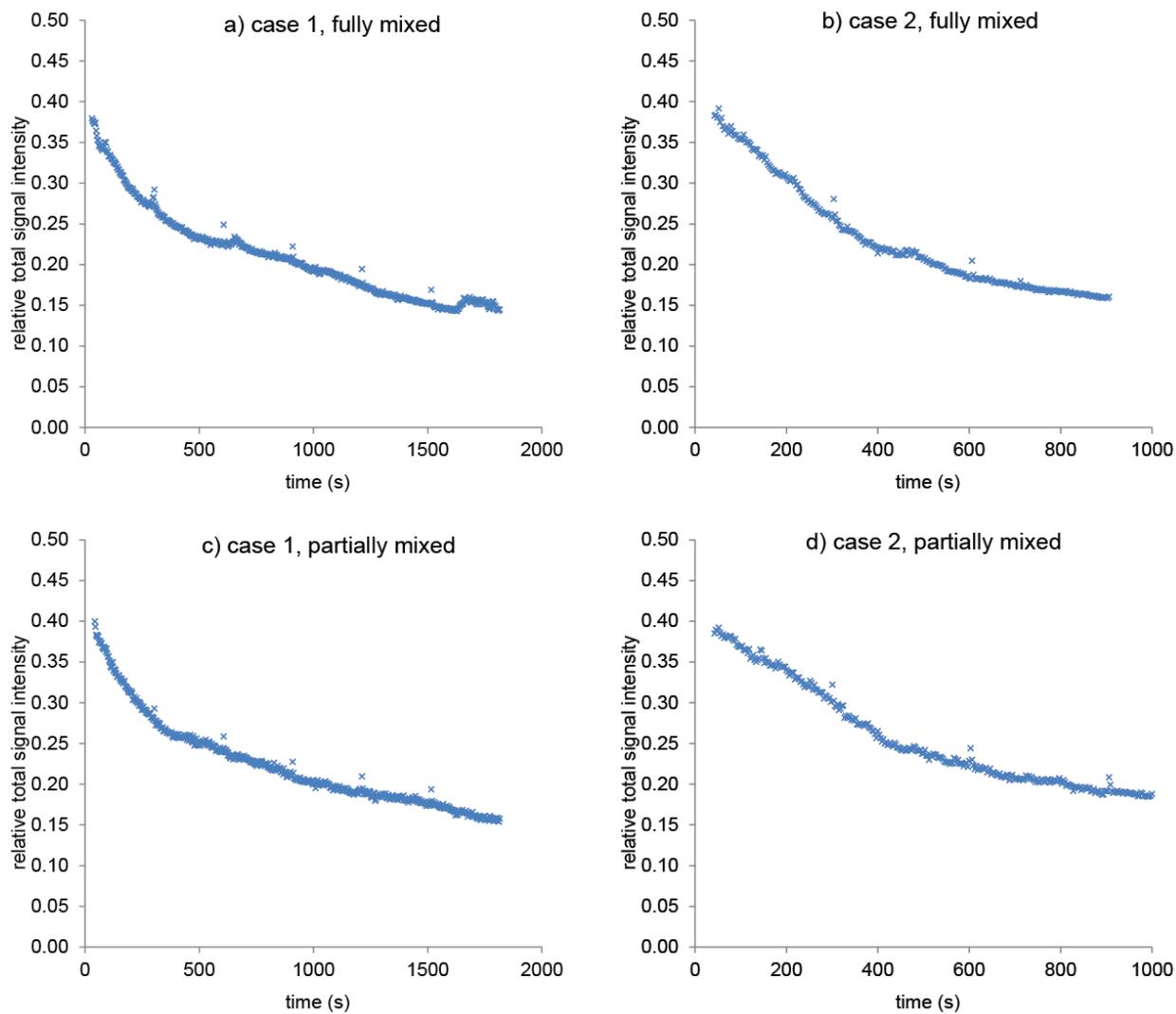

**Fig. 8** Time courses of the relative total signal intensities of anolyte and catholyte in case of heavy water anolyte and light water catholyte (left) and in case of light water anolyte and heavy water catholyte (right)

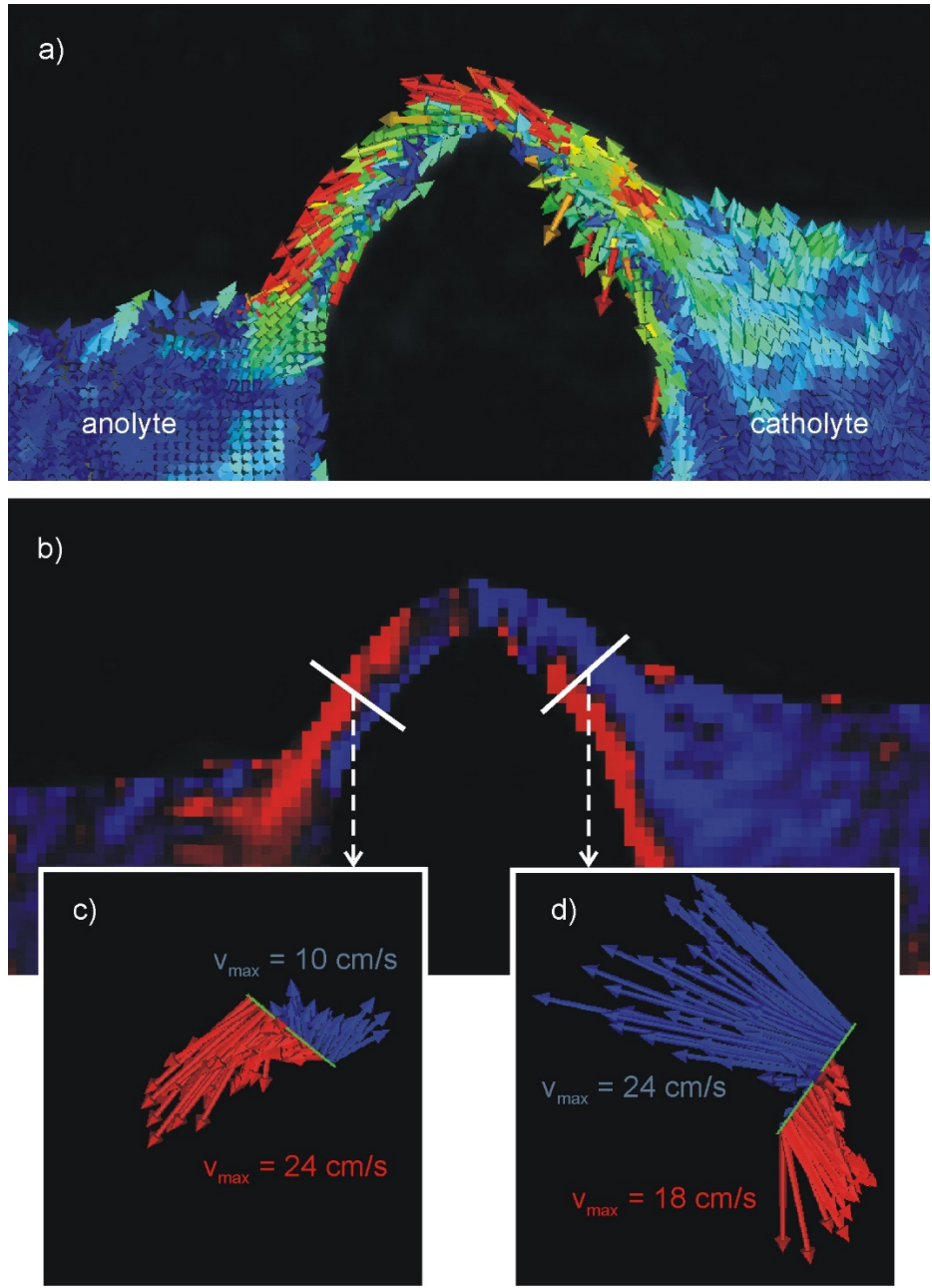

**Fig. 9** Velocity field in a light water bridge determined by tri-directional MR phase contrast imaging. Upper panel (a) displays velocities in the central sagittal slices as color encoded 3D vectors. For better visualization of layer structure of upward and backward flow, the central panel (b) displays only vertical velocity component (red is downward, blue is upward). Multiplanar reformatted cut planes for peak velocity evaluation are indicated as white lines and respective velocity distributions are shown in 3D vector representation in panels (c) and (d)